\documentclass[runningheads,a4paper]{llncs}

\usepackage{amssymb}
\usepackage{graphicx}
\usepackage{cite}
\usepackage{graphicx}
\usepackage{listings}
\usepackage{footnote}
\usepackage{url}
\usepackage{mathtools, cuted}
\usepackage{lipsum}
\usepackage{amsmath}
\usepackage{graphicx}
\usepackage{pdflscape} 
\usepackage{rotating} 
\usepackage{pifont}
\usepackage{fixltx2e}
\usepackage{multirow}
\usepackage{tabularx}
\usepackage{makecell}
\usepackage{textcomp}
\usepackage{multirow}
\usepackage{graphicx}
\usepackage{subfig}
\usepackage{wrapfig,lipsum,booktabs}

\usepackage{algorithm}
\usepackage[noend]{algpseudocode}

\usepackage{color}
\usepackage[table,xcdraw]{xcolor}
\usepackage{url}
\urldef{\mailsa}\path|{guangyuan.piao}@insight-centre.org|
\urldef{\mailsb}\path|{john.breslin}@nuigalway.ie|   
\newcommand{\keywords}[1]{\par\addvspace\baselineskip
\noindent\keywordname\enspace\ignorespaces#1}

\begin{document}

\mainmatter  

\title{Factorization Machines Leveraging Lightweight Linked Open Data-enabled Features for Top-N Recommendations}

\titlerunning{Factorization Machines Leveraging Lightweight Linked Open Data-enabled Features}

%
%
\author{Guangyuan Piao%
\and John G. Breslin}
\authorrunning{}

\institute{Insight Centre for Data Analytics\\
National University of Ireland Galway\\
IDA Business Park, Lower Dangan, Galway, Ireland\\
\mailsa\\
\mailsb\\
}

%
%

\toctitle{Lecture Notes in Computer Science}
\tocauthor{Authors' Instructions}
\maketitle

\begin{abstract}

With the popularity of Linked Open Data (LOD) and the associated rise in freely accessible knowledge that can be accessed via LOD, exploiting LOD for recommender systems has been widely studied based on various approaches such as \emph{graph-based} or using different machine learning models with LOD-enabled features. Many of the previous approaches require construction of an additional graph to run graph-based algorithms or to extract path-based features by combining user-item interactions (e.g., likes, dislikes) and background knowledge from LOD. In this paper, we investigate \emph{Factorization Machines} (FMs) based on particularly \emph{lightweight} LOD-enabled features which can be directly obtained via a public SPARQL Endpoint without any additional effort to construct a graph. Firstly, we aim to study whether using FM with these \emph{lightweight} LOD-enabled features can provide competitive performance compared to a learning-to-rank approach leveraging LOD as well as other well-established approaches such as kNN-item and BPRMF. Secondly, we are interested in finding out to what extent each set of LOD-enabled features contributes to the recommendation performance. Experimental evaluation on a standard dataset shows that our proposed approach using FM with lightweight LOD-enabled features provides the best performance compared to other approaches in terms of five evaluation metrics. In addition, the study of the recommendation performance based on different sets of LOD-enabled features indicate that \emph{property-object lists} and \emph{PageRank scores} of items are useful for improving the performance, and can provide the best performance through using them together for FM. We observe that \emph{subject-property lists} of items does not contribute to the recommendation performance but rather decreases the performance.


\end{abstract}

%
%
\section{Introduction}

\begin{figure}[!h]
	\centering
	\includegraphics[width=0.8\linewidth]{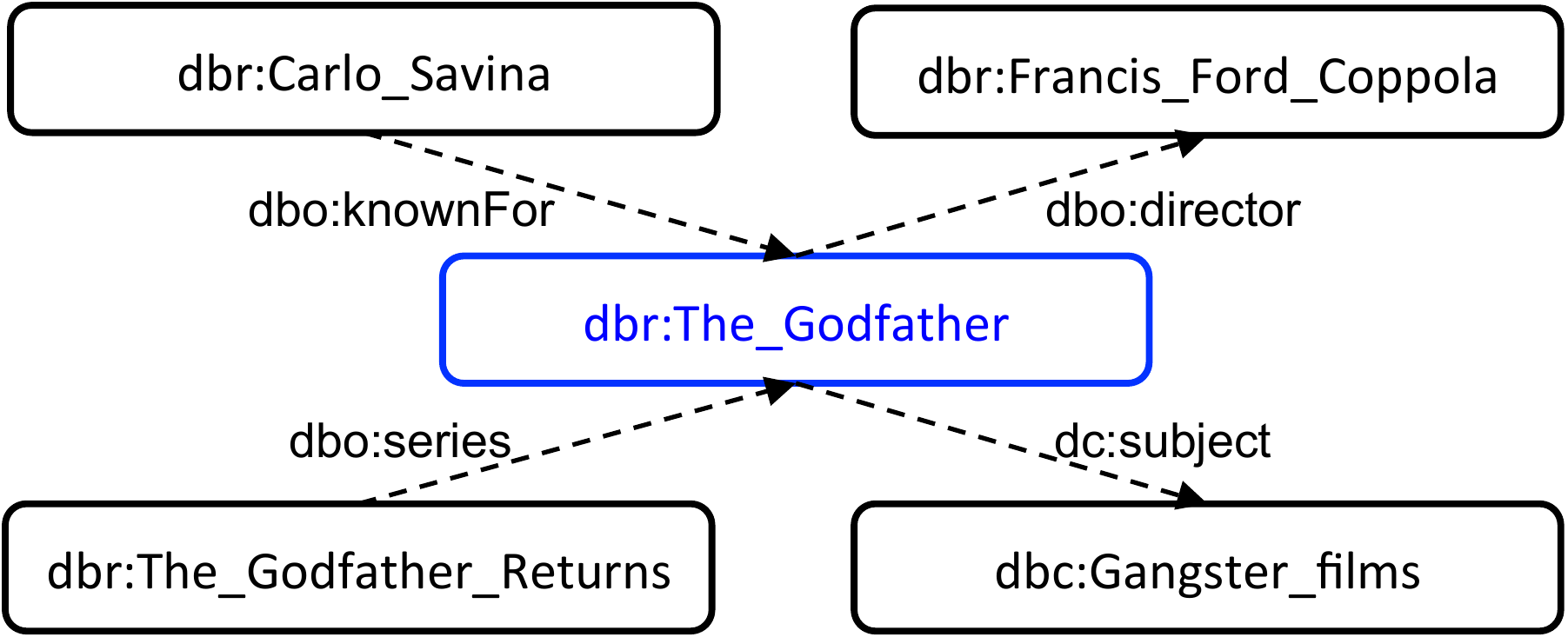}
	\centering
	\caption{An example of background knowledge about the movie \texttt{dbr:The\_Godfather} from DBpedia.}
	\label{example}
\end{figure}

\sloppy {

The term Linked Data, indicates a new generation of technologies responsible for the evolution of the current Web from a Web of interlinked documents to a Web of interlinked data \cite{Heath2011}. Thanks to the Semantic Web's growth and the more recent Linked Open Data (LOD) initiative \cite{Auer2007}, a large amount of RDF\footnote{\url{https://www.w3.org/RDF/}} data has been published in freely accessible datasets. These datasets are connected with each other to form the so-called Linked Open Data cloud\footnote{\url{http://lod-cloud.net/}}. DBpedia \cite{Lehmann2013} which is a 1st-class citizen in this cloud, has become one of the most important and interlinked datasets on the LOD cloud. DBpedia provides cross-domain background knowledge about entities which can be accessible via its SPARQL Endpoint\footnote{\url{http://dbpedia.org/sparql}}. For example, Figure \ref{example} shows pieces of background knowledge about the movie \texttt{dbr\footnote{The prefix \texttt{dbr} denotes for http://dbpedia.org/resource/}:The\_Godfather} in RDF triples, which can be obtained from DBpedia. A RDF triple consists of a subject, a property and an object. As we can see from the figure, there can be incoming knowledge, e.g., \texttt{dbr:Carlo\_Savina}\textrightarrow\texttt{dbo\footnote{The prefix \texttt{dbo} denotes for http://dbpedia.org/ontology/}:knownFor}\textrightarrow \texttt{dbr:The\_Godfather} where \texttt{dbr:The\_Godfather} is used as an object, as well as outgoing knowledge such as \texttt{dbr:The\_Godfather}\textrightarrow\texttt{dbo:director}\textrightarrow\texttt{dbr:Francis\_Ford\_Coppola} where \texttt{dbr:The\_Godfather} is a subject. In the context of the great amount of freely accessible information, many researches have been conducted in order to consume the knowledge provided by LOD for adaptive systems such as recommender systems \cite{DiNoia2014a, DeGemmis2015}. 
}

There have been many approaches for LOD-enabled recommender systems (LODRecSys) such as \emph{semantic similarity/distance} measures, \emph{graph-based} approaches, and \emph{learning-to-rank} approaches by consuming LOD-enabled features. Some previous studies compared their LODRecSys approaches against well-established \emph{collaborative filtering} approaches such as kNN and \emph{matrix factorization} models such as BPRMF \cite{Rendle:2009:BBP:1795114.1795167}, and have shown the benefits of consuming background knowledge powered by LOD. On the other hand, \emph{matrix factorization} models such as BPRMF, which do not exploit LOD-enabled features, have shown competitive performance even compared to some LODRecSys approaches \cite{Musto:2016:SGR:2930238.2930249, Noia2016}. This has in turn motivated us to investigate factorization models consuming LOD-enabled features. 

In this paper, we investigate the use of Factorization Machines (FMs), which can mimic other well-known factorization models such as \emph{matrix factorization}, by leveraging LOD-enabled features. Previous works require increased effort to maintain an additional graph based on user-item interactions and background knowledge about items from LOD in their approaches (We will discuss this in detail in Section 2). In this work, we especially focus on lightweight LOD-enabled features for FM. We define \emph{\textbf{lightweight LOD features}} as features that can be directly obtained via a public SPARQL Endpoint.
\\

The contributions of this work are summarized as follows.
\begin{itemize}
	\item We investigate lightweight LOD-enabled features, which can be directly obtained via the public DBpedia Endpoint, for FM to provide the top-\emph{N} recommendations. Therefore, there is no need to construct a graph which combines user-item interactions (e.g., likes, dislikes) and background knowledge about items. In addition, we investigate to what extent different sets of these features contribute to FM in terms of recommendation performance. \\
	\item We comprehensively evaluate our approach by comparing it to other approaches such as PopRank, kNN, BPRMF, and a state-of-the-art LODRecSys approach SPRank \cite{Noia2016} in terms of five different evaluation metrics.
\end{itemize}

The organization of the rest of the paper is as follows. Section 2 gives some related work, and Section 3 describes our proposed approach using FM with lightweight LOD-enabled features. In Section 4, we describe our experimental setup including the dataset and evaluation metrics. Experimental results are presented in Section 5. Finally, Section 6 concludes the paper with some brief ideas for future work.

%
%
\section{Related Work}
The first attempts to leverage LOD for recommender systems were by \cite{Heitmann2010, Passant2010b}. Heitmann et al. \cite{Heitmann2010} proposed a framework using LOD for open collaborative recommender systems. The Linked Data Semantic Distance (LDSD) measure \cite{Passant2010b} was one of the first works to use LOD for recommender systems in the music domain \cite{Passant2010}. This distance measure considers direct links between two entities/nodes. In addition, it also considers that the same incoming and outgoing nodes via the same properties of two nodes in a graph such as DBpedia. Piao et al. \cite{piao2015A, Guangyuan2016} extended LDSD by investigating different normalization strategies for the paths between two entities. These measures have been designed to work directly on LOD without considering the collaborative view of users. Based on the nature of the graph structure of DBpedia, \emph{graph-based} approaches have been proposed \cite{Musto:2016:SGR:2930238.2930249, Nguyen:2015:ESP:2740908.2742141}. For instance, Musto et al. \cite{Musto:2016:SGR:2930238.2930249} presented a\emph{personalized PageRank} algorithm \cite{Haveliwala2003} using LOD-enabled features for the top-\emph{N} recommendations. Nguyen et al. \cite{Nguyen:2015:ESP:2740908.2742141} investigated \emph{SimRank} \cite{Jeh:2002:SMS:775047.775126} and \emph{PageRank}, and their performance for computing similarity between entities in RDF graphs and investigated their usage to feed a content-based recommender system. Di Noia et al. \cite{DiNoia2012} adapted the Vector Space Model (VSM) to a LOD-based setting, and represented the whole RDF graph as a matrix. On top of the VSM representation, they used the Support Vector Machine (SVM) as a classifier to predict if a user would like an item or not. Using the same representation, they also proposed to assign a weight to each property that represents its worth with respect to the user profile \cite{DiNoia:2012:LOD:2362499.2362501}. In this regard, they used a Genetic Algorithm (GA) to learn the weights of properties that minimize the misclassification errors.
More recently, Di Noia et al. \cite{Noia2016, Ostuni2013} proposed SPRank, which is a semantic path-based approach using learning-to-rank algorithms. This approach first constructed a graph based on user-item interactions and thebackground knowledge of items from LOD. Afterwards, features, called \emph{semantic paths}, were extracted based on the number of paths between a user and an item with min-max normalization. The extracted features were then fed into existing learning-to-rank algorithms such as LMART \cite{Wu2010} provided by RankLib\footnote{\url{https://sourceforge.net/p/lemur/wiki/RankLib/}}. The common requirement for \emph{graph-based} approaches as well as SPRank is that a graph has to be built based on user-item interactions and background knowledge from LOD. Our approach is different as we only consider lightweight LOD-enabled features which can be directly obtained through a public SPARQL Endpoint, and without any additional effort to build a graph. This also makes our model consume updated background knowledge of DBpedia easier when compared to other approaches such as graph-based ones which require downloading a DBpedia dump and building a graph by adding user-item interactions.

There have also been some other interesting directions related to LOD-enabled recommender systems such as the practical LODRecSys \cite{Oliveira2017}, explaining using LOD \cite{Musto:2016:EFE:2959100.2959173}, rating predictions based on matrix factorization with semantic categories \cite{Rowe2014a}, and cross-domain recommendations \cite{Heitmann2014, Heitmann2012}. For example, Oliveira et al. \cite{Oliveira2017} presented a recommender system in the movie domain that consumes LOD (not restricted to DBpedia), which was evaluated with  comparison to seevl (ISWC challenge winner at 2011). Different types of evaluation metrics have been used such as accuracy, novelty etc. The authors from \cite{Musto:2016:EFE:2959100.2959173} presented ExpLOD - a framework which can generate explanations in natural language based on LOD cloud. Musto et al. \cite{Musto:2016:SGR:2930238.2930249} investigated various feature (property) selection strategies and their influences on recommendation performance in terms of accuracy and diversity in movie and book domains. Lalithsena et al. \cite{2793} proposed a novel approach using \emph{type-} and \emph{path-based} methods to extract a subgraph for domain specific recommendation systems. They presented that their approach can decrease 80\% of the graph size without losing accuracy in the context of recommendation systems in movie and book domains. These, although interesting, are however beyond the scope of this paper and we aim to explore them in future work.

%
%
\section{Proposed Method}
In this section, we first briefly introduce FMs and the optimization criteria we used in this study (Section 3.1). Next, we will describe our features from user-item interactions as well as background knowledge from DBpedia (Section 3.2).

\subsection{Factorization Machines}
Factorization Machines (FMs) \cite{Rendle2010a}, which can mimic other well known factorization models such as \emph{matrix factorization}, \emph{SVD++} \cite{Koren:2009:CFT:1557019.1557072}, have been widely used for collaborative filtering tasks \cite{Rendle:2012:FML:2168752.2168771}. FMs are able to incorporate the high-prediction accuracy of factorization models and flexible feature engineering. An important advantage of FMs is the model equation 

\begin{equation}
	\hat{y}^{FM}(x) = w_0 + \sum_{i=1}^pw_ix_i+\sum_{i=1}^{p}\sum_{j>i}^{p}<v_i,v_j>x_ix_j
\end{equation}
\\
where $w_0 \in \mathbb{R}, x\ and\ w \in \mathbb{R}^p, v_i \in \mathbb{R}^m $. The first part of the FM model captures the interactions of each input variable $x_i$, while the second part of it models all pairwise interactions of input variables $x_ix_j$. Each variable $x_i$ has a latent factor $v_i$, which is a \emph{m}-dimensional vector allows FMs work well even in highly sparse data. 

\subsubsection{Optimization.}
In this work, we use a \emph{pairwise} optimization approach - Bayesian Personalized Ranking (BPR). The loss function was proposed by Rendle et al. \cite{Rendle:2009:BBP:1795114.1795167}.

\begin{equation}
	l(x_1, x_2) = \sum_{x_1 \in C_u^+} \sum_{x_2 \in C_u^-} (-\log[\delta(\hat{y}^{FM}(x_1) - \hat{y}^{FM}(x_2))])  
\end{equation}
\\
where $\delta$ is a sigmoid function: $\delta(x)=\frac{1}{1+e^{-x}}$, and $C_u^+$ and $C_u^-$ denote the set of positive and negative feedbacks respectively. L2-regularization is used for the loss function.

\subsubsection{Learning.} We use the well-known \emph{stochastic gradient descent} algorithm to learn the parameters in our model. To avoid overfitting on the training dataset, we adopt an early stopping strategy as follows.

\begin{enumerate}
	\item Split the dataset into training and validation sets.
	\item Measure the current loss on the validation set at the end of each epoch.
	\item Stop and remember the epoch if the loss has increased.
	\item Re-train the model using the whole dataset.	
\end{enumerate}

\subsection{Features}

\begin{figure}[!t]
	\centering
	\includegraphics[width=0.8\linewidth]{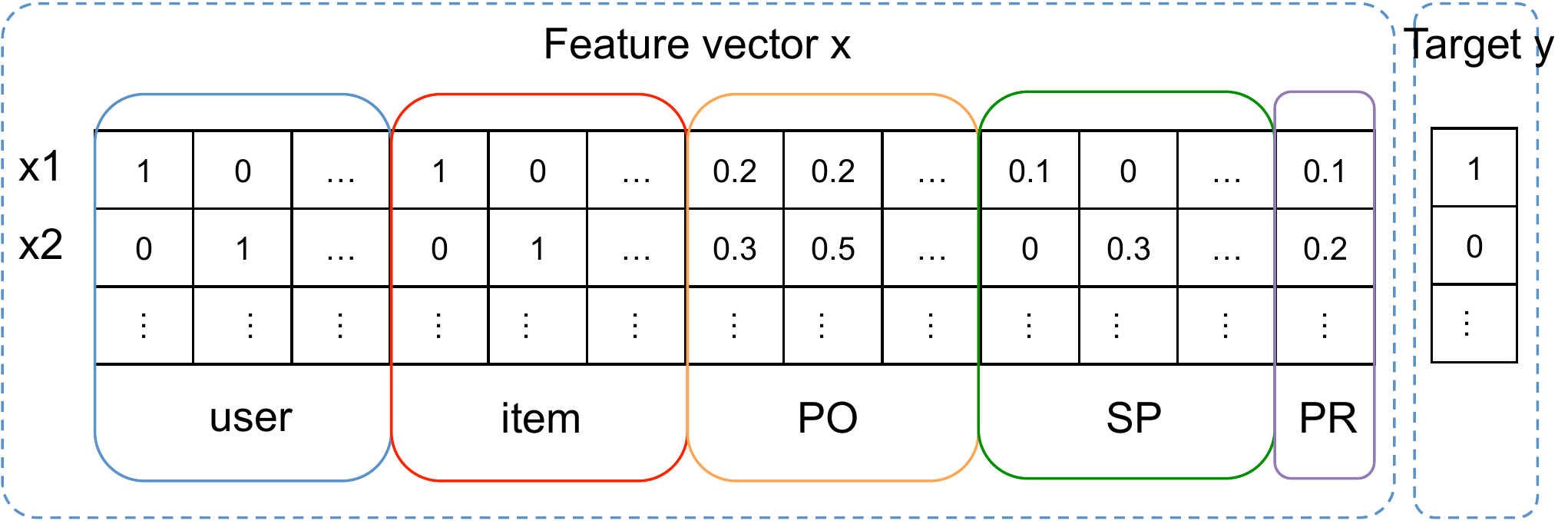}
	\centering
	\caption{Overview of features for Factorization Machine. PO denotes all property-objects, and SP denotes all subject-property for items in the dataset. PR denotes the PageRank scores of items.}
	\label{fig:overview}
\end{figure}

Figure \ref{fig:overview} presents the overview of features for our FM. The details of each set of features are described below.

\subsubsection{User and item index.} The first two sets of features indicate the indexes of the user and item in a training example. A feature value equals 1 for the corresponding user/item index, e.g., $val(U_i)=1$ and $val(I_j)=1$ denote an example about the \emph{i}-th user and \emph{j}-th item.

\subsubsection{Property-Object list (PO).} This set of features denotes all property-objects of an item \emph{i} when \emph{i} is a subject in RDF triples. This set of features can be obtained easily by using a SPARQL query as shown below via the DBpedia SPARQL Endpoint.

\begin{verbatim}
PREFIX dbo:<http://dbpedia.org/ontology/>
PREFIX dct:<http://purl.org/dc/terms/>

SELECT DISTINCT ?p ?o WHERE { { <itemURI> ?p ?o  .
FILTER REGEX(STR(?p), ``^http://dbpedia.org/ontology'') . 
FILTER (STR(?p) NOT IN (dbo:wikiPageRedirects, 
dbo:wikiPageExternalLink)) . FILTER ISURI(?o) } 
UNION { <itemURI> ?p ?o . FILTER ( STR(?p) IN (dct:subject) ) } }
\end{verbatim}

\begin{figure}[!h]
	\centering
	\includegraphics[width=\linewidth]{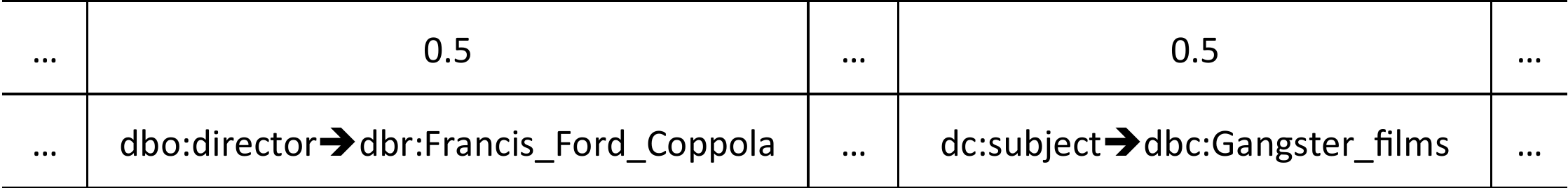}
	\centering
	\caption{An example for PO values for the movie \texttt{dbr:The\_Godfather} in Figure \ref{example}.}
	\label{POex}
\end{figure}

An intuitive way of giving feature values for PO might be to assign 1 for all property-objects of an item \emph{i} ($PO_i$). However, it can be biased as some entities in DBpedia have a great number of property-objects while others do not. Therefore, we normalize the feature values of $PO_i$ based on the size of $PO_i$ so that all the feature values of $PO_i$ sum up to 1. Formally, the feature value of \emph{j}-th property-object for an item \emph{i} is measured as $val(PO_{i}(j))=\frac{1}{|PO_i|}$. Take the graph in Figure \ref{example} as an example, as we have two property-objects for the movie \texttt{dbr:The\_Godfather}, each property-object of the movie will have a feature value of 0.5, respectively (see Figure \ref{POex}).

\subsubsection{Subject-Property list (SP).} Similar to the PO, we can obtain incoming background knowledge about an item \emph{i} where \emph{i} is an object in RDF triples. This set of features can be obtained by using a SPARQL query as shown below.

\begin{verbatim}
PREFIX dbo:<http://dbpedia.org/ontology/>

SELECT DISTINCT ?s ?p WHERE { ?s ?p <itemURI> .
FILTER REGEX(STR(?p), ``^http://dbpedia.org/ontology'') . 
FILTER (STR(?p) NOT IN (dbo:wikiPageRedirects, 
dbo:wikiPageExternalLink, dbo:wikiPageDisambiguates) } 
\end{verbatim}

In the same way as we normalized feature values of $PO_i$ for an item \emph{i}, we normalize the feature values of $SP_i$ based on the size of $SP_i$ so that all the feature values of $SP_i$ sum up to 1. The feature value of the \emph{j}-th SP for an item \emph{i} is measured as $val(SP_{i}(j))=\frac{1}{|SP_i|}$.

\subsubsection{PageRank score (PR).} PageRank \cite{Page1999} is a popular algorithm with the purpose of measuring the relative importance of a node in a graph. In order to capture the importance of an entity in Wikipedia/DBpedia, Thalhammer et al. \cite{Thalhammer2016} proposed providing PageRank scores of all DBpedia entities, which are based on links using \texttt{dbo:wikiPageWikiLink} among entities. A PageRank score of an item (entity) might be a good indicator of the importance of an entity for recommendations in our case. The PageRank score of a DBpedia entity can be obtained by using the SPARQL as shown below.

\begin{verbatim}
PREFIX rdf:<http://www.w3.org/1999/02/22-rdf-syntax-ns#>
PREFIX dbo:<http://dbpedia.org/ontology/>
PREFIX vrank:<http://purl.org/voc/vrank#>

SELECT ?score FROM <http://dbpedia.org> 
FROM <http://people.aifb.kit.edu/ath/#DBpedia_PageRank>
WHERE { <itemURI> vrank:hasRank/vrank:rankValue ?score . }
\end{verbatim}

The scale of PageRank scores is different from other feature values, which can delay the convergence of learning parameters for our model. In this regard, we normalize the PageRank scores by their maximum value.

\begin{equation}
	val(PR_i) = \frac{PageRank_i}{max(PageRank_j, j \in I)}
\end{equation}
\\
where $PageRank_i$ denotes the original PageRank score of \emph{i} which is obtained from the SPARQL Endpoint, and $max(PageRank_j, j \in I)$ denotes the maximum PageRank score of all items.

%
%

\section{Experimental Setup}

In this section, we introduce the dataset for our experiment (Section 4.1) and five evaluation metrics for evaluating the performance of the recommendations (Section 4.2). Afterwards, we describe four methods that have been used for comparison with our approach for evaluation (Section 4.3).

\subsection{Dataset}

We used the Movielens dataset from \cite{Noia2016}. The dataset was originally from the Movielens dataset\footnote{\url{https://grouplens.org/datasets/movielens/1m/}}, which consists of users and their ratings about movie items. To facilitate LODRecSys, each of the items in this dataset has been mapped into DBpedia entities if there is a mapping available\footnote{\url{http://sisinflab.poliba.it/semanticweb/lod/recsys/datasets/}}. In the same way as \cite{Noia2016}, we consider ratings higher than 3 as positive feedback and others as negative one. Table \ref{tb:dataset} shows details about the dataset. The dataset consists of 3,997 users and 3,082 items with 695,842 ratings where 56\% of them are positive ratings. We split the dataset into training (80\%) and test (20\%) sets for our experiment.
\newcolumntype{C}{>{\centering\arraybackslash}p{14em}}
\begin{table}[!h]
\renewcommand{\arraystretch}{1.2}
\centering
\caption{Movielens dataset statistics}
\label{tb:dataset}
\begin{tabular}{C|C}
\Xhline{3\arrayrulewidth}
\# of users            & 3,997                         \\ \hline
\# of items            & 3,082                            \\ \hline
\# of ratings          & 695,842                   \\ \hline
avg. \# of ratings     & 174                        \\ \hline
sparsity               & 94.35\%                           \\ \hline
\% of positive ratings & 56\%                        \\  \Xhline{3\arrayrulewidth}
\end{tabular}
\end{table}

\subsection{Evaluation Metrics}

We use five different evaluation metrics to measure the quality of recommendations provided by different approaches.

\begin{itemize}
	\item \textbf{P@N}: The Precision at rank \emph{N} represents the mean probability that retrieved items within the top-\emph{N} recommendations are relevant to the user.
	
	\begin{equation}
	P@N = \frac{|\{relevant\ items\}| \cap |\{retrieved\ items@n\}|}{|\{retrieved\ items\}|}
	\end{equation}
	
	\item \textbf{R@N}: The Recall at rank \emph{N} represents the mean probability that relevant items are successfully retrieved within the top-\emph{N} recommendations.
	
	\begin{equation}
	R@N = \frac{|\{relevant\ items\}| \cap |\{retrieved\ items@n\}|}{|\{relevant\ items\}|}
	\end{equation}
	
	\item \textbf{nDCG@N}: Precision and recall consider the relevance of items only. On the other hand, nDCG takes into account the relevance of items as well as their rank positions.
	
	\begin{equation}
	nDCG@N = \frac{1}{IDCG@N} \sum_{k=1}^{N} \frac{2^{\hat{r}_{uk}}-1}{\log_2(1+k)}
	\end{equation}
	Here, $\hat{r}_{uk}$ denotes the rating given by a user \emph{u} to the item in position \emph{k} in the top-\emph{N} recommendations, and IDCG@\emph{N} denotes the score obtained by an ideal or perfect top-\emph{N} ranking and acts as a normalization factor.
	\item \textbf{MRR}: The Mean Reciprocal Rank (MRR) indicates at which rank the first link \emph{relevant} to the user occurs (denoted by $rank_k$) on average. 
	
	\begin{equation}
	MRR = \frac{1}{|U|}\sum_{k=1}^{|U|}\frac{1}{rank_k}
	\end{equation}
	
	\item \textbf{MAP}: The Mean Average Precision (MAP) measures the average of the average precision (AP) of all liked items for all users. For each user, the average precision of the user is defined as:
	
	\begin{equation}
		AP = \frac{\sum_{n=1}^{N}P@n \times like(n)}{|I|}
	\end{equation}
	where \emph{n} is the number of items, $|I|$ is the number of liked items of the user, and $like(n)$ is a binary function to indicate whether the user prefers the \emph{n}-th item or not.

\end{itemize}

The \emph{bootstrapped paired t-test}, which is an alternative to the paired t-test when the assumption of normality of the method is in doubt, is used for testing the significance where the significance level was set to 0.01 unless otherwise noted. 

\subsection{Compared Methods}

We use four approaches including a baseline PopRank and other methods which have been frequently used in the literature \cite{Musto:2016:SGR:2930238.2930249, Noia2016} to evaluate our proposed method.

\begin{itemize}
	
	\item \textbf{PopRank:} This is a non-personalized baseline approach which recommends items based on the popularity of each item. 
	\\
	\item \textbf{kNN-item:} This is a collaborative filtering approach based on the \emph{k} most similar items. We use a MyMedialiite \cite{Gantner:2011:MFR:2043932.2043989} implementation for this baseline where $k=80$.
	\\
	\item \textbf{BPRMF \cite{Rendle:2009:BBP:1795114.1795167}:} This is a matrix factorization approach for learning latent factors for users and items. We use a MyMedialiite \cite{Gantner:2011:MFR:2043932.2043989} implementation for this baseline where the dimensionality of the factorization $m=200$.
	\\
	\item \textbf{SPRank \cite{Noia2016}:} This is a \emph{learning-to-rank} approach for LODRecSys based on \emph{semantic paths} extracted from a graph including user-item interactions (e.g., likes, dislikes, etc.) as well as the background knowledge obtained from DBpedia. In detail, \emph{semantic paths} are sequences of properties including \emph{likes} and \emph{dislikes} based on user-item interactions. For example, given the graph information \texttt{user1}\textrightarrow\texttt{likes}\textrightarrow\texttt{item1}\textrightarrow\texttt{p1}\textrightarrow\texttt{item2}, a semantic path (\texttt{likes}, \texttt{p1}) can be extracted from \texttt{user1} to \texttt{item2}. Another difference between SPRank \cite{Noia2016} and our approach in terms of features is that the authors considered property-objects for each item including the property \texttt{dbo:wikiPageWikiLink} which cannot be queried via the DBpedia Endpoint but requires settings up a local endpoint using a DBpedia dump. On the other hand, we only considers sets of LOD-enabled features which can be obtained from a public DBpedia Endpoint. We use LMART \cite{Wu2010} as the learning algorithm for SPRank as this approach overall provides the best performance compared to other learning-to-rank algorithms in \cite{Noia2016}. We used the author's implementation\footnote{\url{https://github.com/sisinflab/lodreclib}} which has been optimized for nDCG@10.
\end{itemize}

%
%
\section{Results}

In this section, we first compare our approach to the aforementioned methods in terms of five evaluation metrics (Section 5.1). We denote our approach as LODFM, and the results of LODFM are based on best tuned parameters, i.e., $m=200$ using PO and PR as LOD-enabled features. We discuss self comparison by using different sets of features, as well as different dimensionality \emph{m} for factorization, in detail in Section 5.2.

\newcolumntype{C}{>{\centering\arraybackslash}p{5.5em}}
\begin{table}[!t]
\renewcommand{\arraystretch}{1.3}
\centering
\caption{Recommendation performance compared to baselines in terms of five different evaluation metrics. The best performing strategy is in bold.}
\label{comparison}
\begin{tabular}{|C|C|C|C|C|C|}
\hline
                 & \textbf{PopRank} & \textbf{kNN-item} & \textbf{BPRMF} & \textbf{SPRank} & \textbf{LODFM} \\ \hline \hline
\textbf{MRR}     & 0.4080           & 0.5756       & 0.5906         & 0.3013          & \textbf{0.6218}      \\ \hline
\textbf{MAP}     & 0.1115           & 0.2037       & 0.2018         & 0.0612          & \textbf{0.2318} \\ \hline
\textbf{nDCG@1}  & 0.2459           & 0.4086       & 0.4269         & 0.1758          & \textbf{0.4685} \\ \hline
\textbf{P@1}     & 0.2459           & 0.4086       & 0.4269         & 0.1758          & \textbf{0.4685} \\ \hline
\textbf{R@1}     & 0.0064           & 0.0132       & 0.0258         & 0.0082          & \textbf{0.0268} \\ \hline
\textbf{nDCG@5}  & 0.2809           & 0.4049       & 0.4176         & 0.2195          & \textbf{0.4537} \\ \hline
\textbf{P@5}     & 0.2240           & 0.3538       & 0.3393         & 0.1287          & \textbf{0.3829} \\ \hline
\textbf{R@5}     & 0.0305           & 0.0553       & 0.0977         & 0.0291          & \textbf{0.1052}          \\ \hline
\textbf{nDCG@10} & 0.3664           & 0.4753       & 0.5000         & 0.2845          & \textbf{0.5231} \\ \hline
\textbf{P@10}    & 0.2104           & 0.3179       & 0.2883         & 0.1068          & \textbf{0.3256} \\ \hline
\textbf{R@10}    & 0.0580           & 0.0978       & 0.1602         & 0.0488          & \textbf{0.1730}          \\ \hline
\end{tabular}
\end{table}

\subsection{Comparison with Baselines}
The results of comparing our proposed approach with the baselines are presented in Table \ref{comparison} in terms of MRR, MAP, nDCG@\emph{N}, P@\emph{N} and R@\emph{N}.

Overall, LODFM provides the best performance in terms of all evaluation metrics. In line with the results from \cite{Noia2016}, SPRank does not perform as well on the Movielens dataset compared to other collaborative filtering approaches such as kNN and BPRMF. On the other hand, we observe that LODFM significantly outperforms SPRank as well as other baseline methods. Among baselines, kNN-item is the best performing method in terms of P@5 and P@10 while BPRMF is the best performing baseline in terms of other evaluation metrics. A significant improvement of LODFM over BPRMF in MRR (+5.3\%), MAP (+14.9\%), nDCG@10 (+4.6\%), P@10 (+12.9\%) and R@10 (+8\%) can be noticed. The results indicate that LOD-enabled features are able to improve the recommendation performance for factorization models. Compared to kNN-item, LODFM improves the performance by 8.2\% and 2.4\% in terms of P@5 and P@10, respectively. It is also interesting to observe that factorization models such as BPRMF and LODFM have much better performance especially in terms of recall compared to kNN-item. For example, LODFM improves the performance by 103\%, 90\% and 76.9\% in terms of recall when $N=$ 1,5 and 10, respectively.

\newcolumntype{C}{>{\centering\arraybackslash}p{6.5em}}
\begin{table}[!t]
\renewcommand{\arraystretch}{1.3}
\centering
\caption{Recommendation performance of LODFM using different sets of features such as property-object list (PO), subject-property list (SP) and PageRank scores (PR). The best performing strategy is in bold.}
\label{tb:featureCompare}
\begin{tabular}{|C|C|C|C|C|}
\hline

                 & \textbf{PO} & \textbf{PO+SP} & \textbf{PO+PR} & \textbf{PO+SP+PR}  \\ \hline \hline
\textbf{MRR}     & 0.5769           & 0.5403       & \textbf{0.5783}         & 0.5561               \\ \hline
\textbf{MAP}     & \textbf{0.2096}           & 0.1957       & 0.2080         & 0.2008          \\ \hline
\textbf{nDCG@1}  & 0.4224           & 0.3788       & \textbf{0.4236}         & 0.3971           \\ \hline
\textbf{P@1}     & 0.4224           & 0.3788       & \textbf{0.4236}         & 0.3971           \\ \hline
\textbf{R@1}     & 0.0237           & 0.0210       & \textbf{0.0241}         & 0.0223         \\ \hline
\textbf{nDCG@5}  & 0.4152           & 0.3861       & \textbf{0.4214}         & 0.3963           \\ \hline
\textbf{P@5}     & 0.3459           & 0.3222       & \textbf{0.3479}         & 0.3280         \\ \hline
\textbf{R@5}     & 0.0931           & 0.0841       & \textbf{0.0934}         & 0.0866                 \\ \hline
\textbf{nDCG@10} & 0.4904           & 0.4627       & \textbf{0.4945}         & 0.4743         \\ \hline
\textbf{P@10}    & 0.2973           & 0.2805       & \textbf{0.2975}         & 0.2860       \\ \hline
\textbf{R@10}    & \textbf{0.1558}           & 0.1436       & 0.1541         & 0.1476                    \\ \hline
\end{tabular}
\end{table}

\subsection{Model Analysis}

\subsubsection{Analysis of features.} To better understand the contributions of each feature set for recommendations, we discuss the recommendation performance with different sets of features for FM in this section. Table \ref{tb:featureCompare} shows the recommendation performance of LODFM using different features with $m=10$. The two fundamental features - user and item indexes are included by default and omitted from the table for clarity. 

Overall, using a property-object list (PO) and the PageRank score (PR) of items provides the best performance compared to other strategies. As we can see from Table \ref{tb:featureCompare}, PO+PR improves the recommendation performance compared to PO in terms of most of the evaluation metrics. Similar results can be observed by comparing PO+SP+PR against PO+SP, which shows the importance of PageRank scores of items. On the other hand, the performance is decreased by including SP, e.g., PO+SP vs. PO and PO+SP+PR vs. PO+PR. This shows that incoming knowledge about movie items is not helpful in improving recommendation performance.

\subsubsection{Analysis of dimensionality \emph{m} for factorization.} 

\begin{figure}[!t]
  \centering
  \subfloat[MRR (Mean Reciprocal Rank)]{\includegraphics[width=0.5\textwidth]{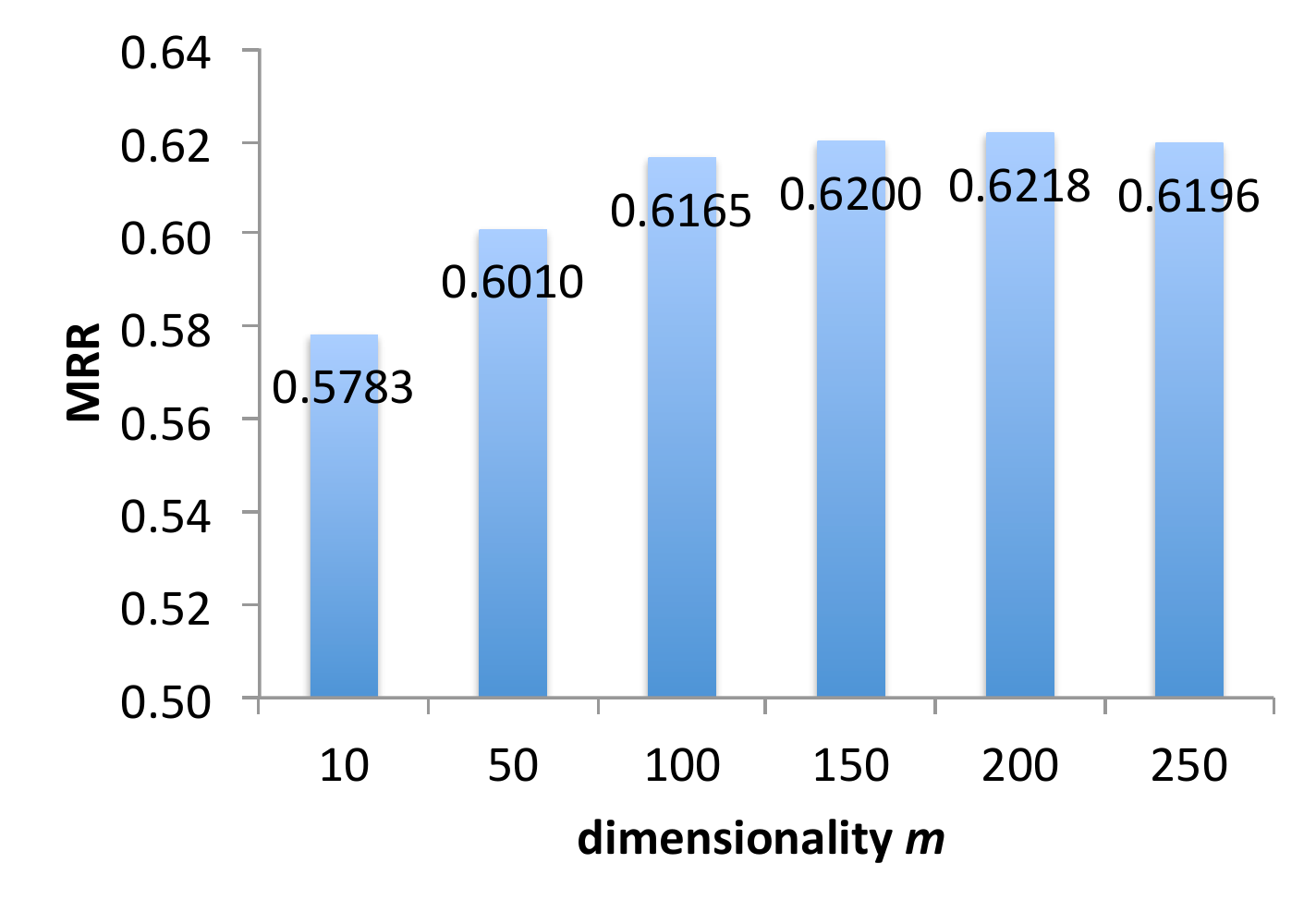}\label{fig:f1}}
  \hfill
  \subfloat[nDCG@N and P@N]{\includegraphics[width=0.5\textwidth]{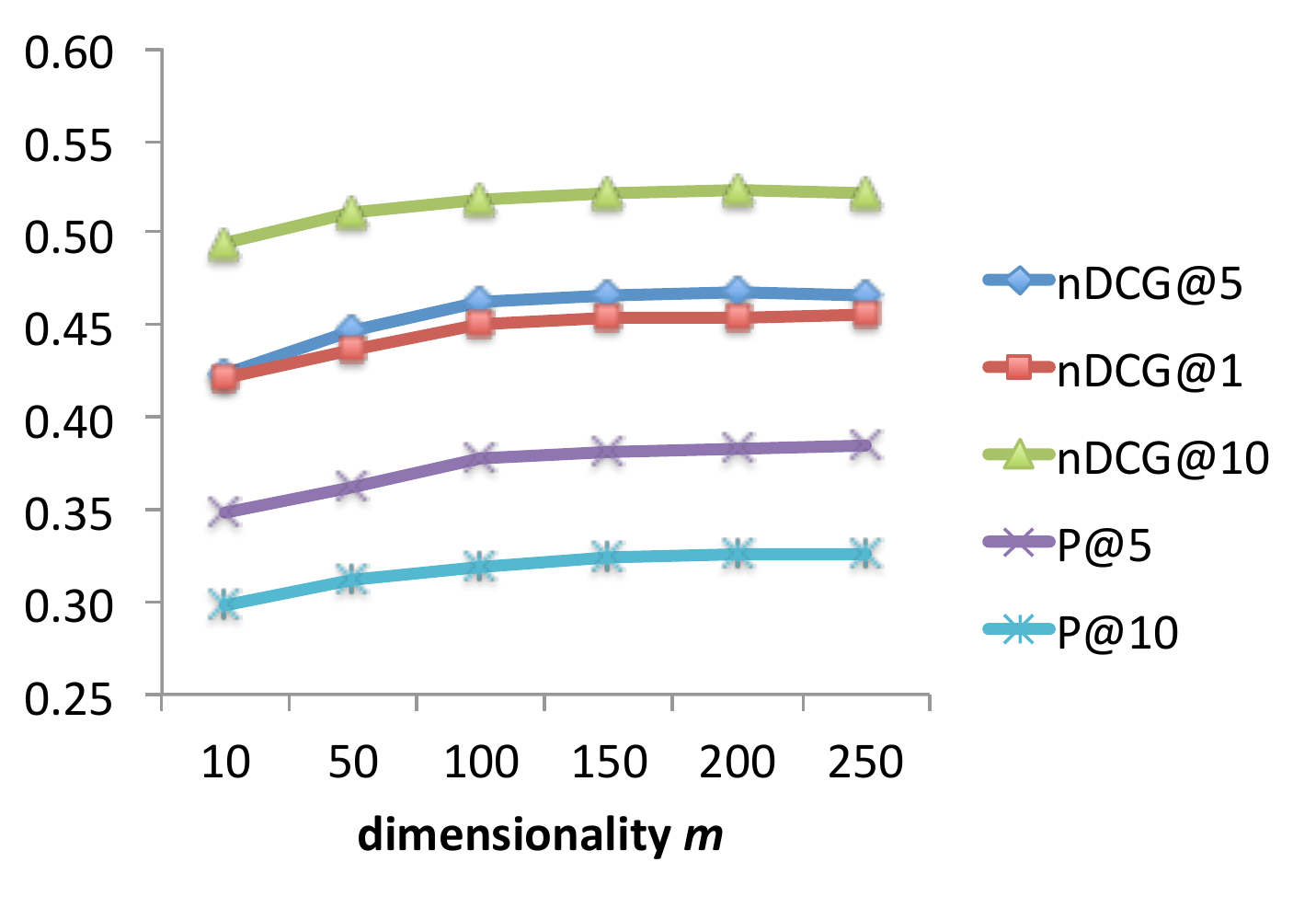}\label{fig:f3}}
  \hfill
  \subfloat[MAP (Mean Average Precision)]{\includegraphics[width=0.5\textwidth]{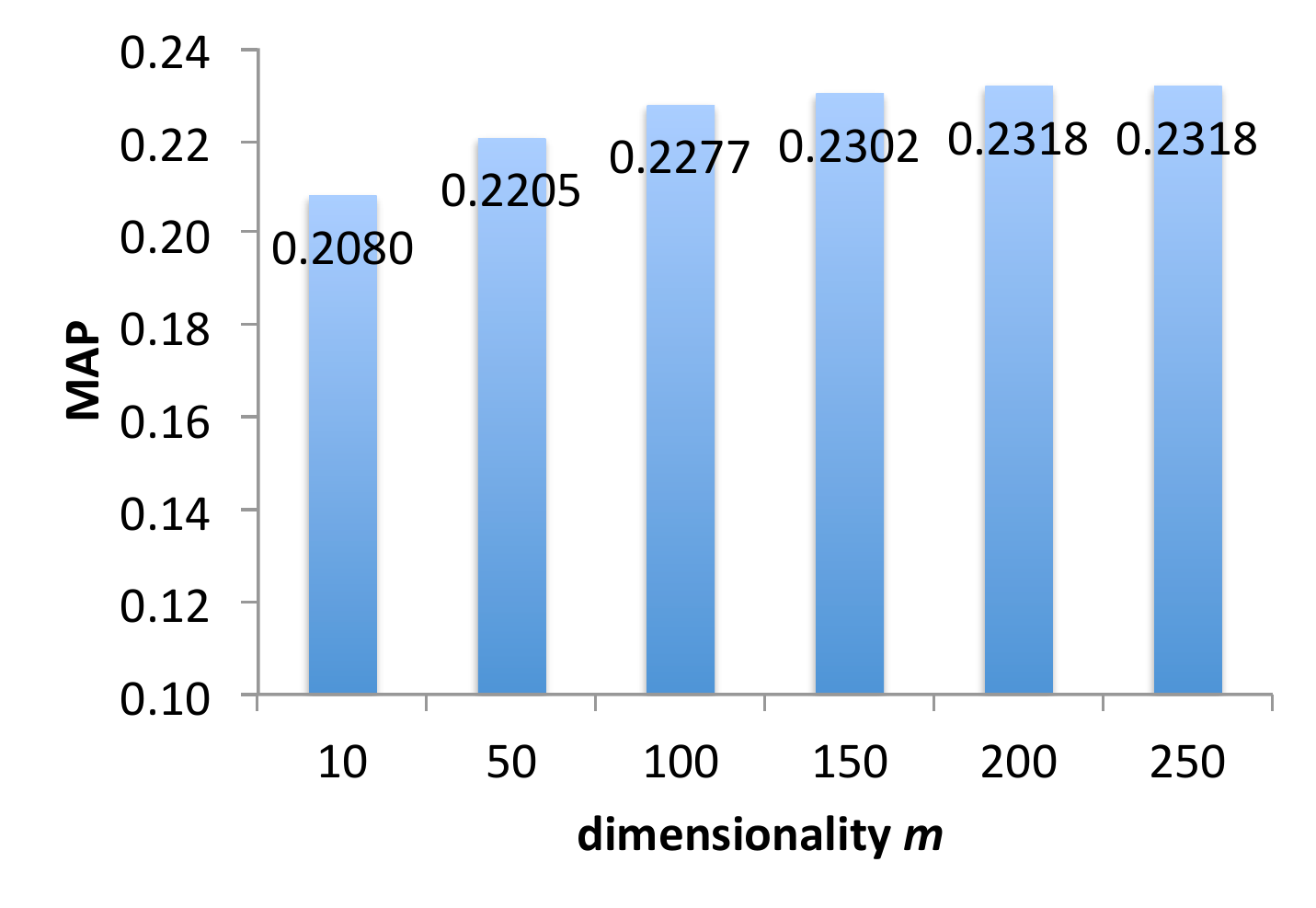}\label{fig:f2}}
  \hfill
  \subfloat[Recall@N]{\includegraphics[width=0.5\textwidth]{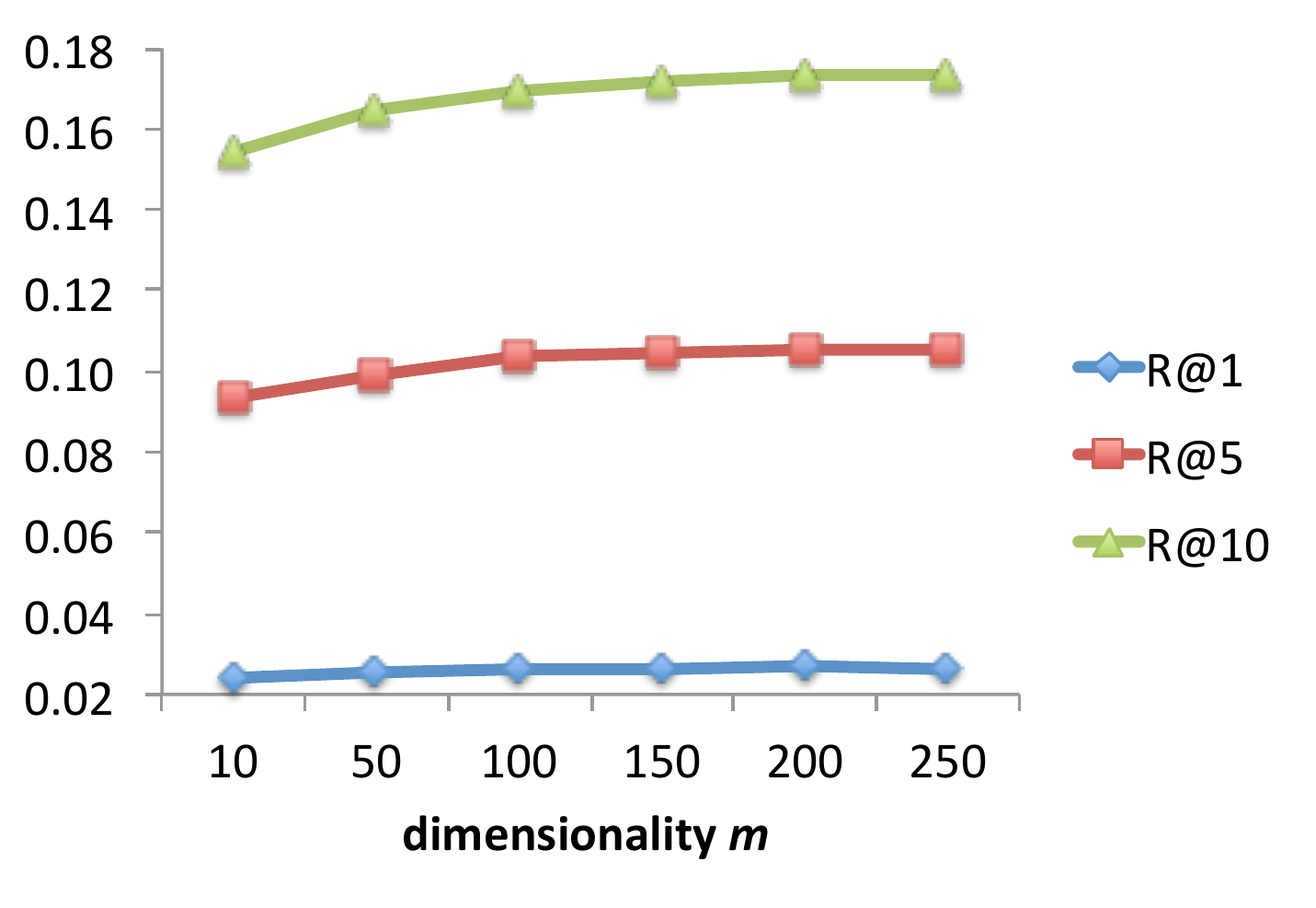}\label{fig:f4}}
  \caption{Recommendation performance based on different values for the dimensionality \emph{m} of FM using PO+PR in terms of different evaluation metrics.}
  \label{dim}
\end{figure}

The dimensionality of factorization plays an important role in capturing pairwise interactions of input variables when \emph{m} is chosen large enough \cite{Rendle:2012:FML:2168752.2168771}. Figure \ref{dim} illustrates the recommendation performance using different values for the dimensionality of factorization (The results of P@1 are equal to nDCG@1 and therefore omitted from Figure \ref{dim}(b)) using PO and PR as LOD-enabled features. As we can see from the figure, the performance consistently increases with higher values of \emph{m} until $m=200$ in terms of five evaluation metrics. For example, the performance is improved by 7.5\% and 11.4\% in terms of MRR and MAP with $m=200$ compared to $m=10$. There is no significant improvement with values higher than 200 for \emph{m}.

%
%
\section{Conclusions}

In this paper, we investigated using FM with lightweight LOD-enabled features, such as property-object lists, subject-property lists, and PageRank scores of items which can be directly obtained from the DBpedia SPARQL Endpoint, for top-\emph{N} recommendations. The results show that our proposed approach significantly outperforms compared approaches such as SPRank, BPRMF. In addition, we analyzed the recommendation performance based on different combinations of features. The results indicate that using the property-object list and the PageRank scores of items can provide the best performance. On the other hand, including the subject-property list of items is not helpful in improving the quality of recommendations but rather decreases the performance. In the future, we plan to evaluate our approach using other datasets in different domains. Furthermore, we aim to investigate other lightweight LOD-enabled features which might be useful to improve the recommendation performance.


\bibliographystyle{splncs03} 
\bibliography{library}

\end{document}